\documentclass[prb,twocolumn,superscriptaddress]{revtex4}
\usepackage{amsfonts}
\usepackage{amsmath}
\usepackage{amssymb}
\usepackage{graphicx}
\usepackage{hyperref}
\usepackage{bbold}

\begin{document}

\title{ Investigation into the inadequacy of cRPA in reproducing screening in strongly correlated systems }
\author{Qiang Han}
\affiliation{Department of Physics \& Astronomy, Rutgers University, Piscataway, NJ 08854-8019, USA}
\author{B. Chakrabarti}
\affiliation{Department of Physics \& Astronomy, Rutgers University, Piscataway, NJ 08854-8019, USA}
\author{K. Haule}
\affiliation{Department of Physics \& Astronomy, Rutgers University, Piscataway, NJ 08854-8019, USA}

\date{\today}

\begin{abstract}
The accuracy of the constrained random phase approximation(cRPA) method is examined in multi-orbital Hubbard models containing all possible on-site density-density interactions. Using DMFT, we show that the effective model constructed using cRPA fails to reproduce the spectral properties of the original full model in a wide parameter range.  By comparing quantities such as the density of states and quasiparticle residues of the full and the effective models, we show that cRPA systematically overestimates the screening of Hubbard U for DMFT impurity solvers. We instead propose a new method to investigate the screening mechanism in the system using the local polarization, which is highly successful in reproducing spectra  and also shows that the true screening is far less than that predicted by RPA. Furthermore, we compare the fully screened interaction $W$ given by RPA and our new method and show that the RPA $W$ is also overscreened and misses the signatures of local screening, which are clearly present in our new method.
\end{abstract}
\maketitle

\section{Introduction}
 Achieving a truly \emph{ab initio} description of complex correlated materials is one of the prime objectives of condensed matter physics today. These compounds attract lots of interest as they often have very complicated phase diagrams displaying a variety of interesting phenomena such as  metal-Mott insulating transition(MIT), unconventional superconductivity and non-trivial magnetic order,charge/spin density waves etc\cite{RMP_MIT_1998_Imada}. These phenomena cannot be explained by free electron-based approximations and often lie beyond the scope of density functional theory(DFT), the workhorse for predicting properties of solids from first principles\cite{RMP_dft_review_1989_R_O_Jones}. Dynamical Mean Field Theory (DMFT) seeks to overcome some of the difficulties of studying these systems due to strong correlations between electrons by mapping the lattice problem to a numerically tractable auxiliary impurity problem coupled to a bath which is determined self-consistently. This approach, which was originally created to study model hamiltonians such as the Hubbard model, has recently been combined with DFT (DFT+DMFT) and has proved to be highly successful in explaining the properties of strongly correlated materials \cite{RMP_LDA+DMFT_2006_G.K}. Various implementations of DFT+DMFT are currently available\cite{PRB_Wannier_Downfolding_2006_D.Vollhardt,PRB_Wannier_Downfolding_2006_O.K.Andersen,PRB_localorbital_Downfolding_2008_Lichtenstein,PRB_dmft_wien2k_2010_Chuck_Haule,PRB_LaFeAsO_2010_A.Georges,PRB_covalency_TMoxides_2014_k.Haule}, which mainly differ in i)the choice of how to project to the localized impurity degrees of freedom and ii)the energy window used while embedding the impurity self-energy into the DFT lattice eigensystem. Though the relative merits of a particular scheme might be dependent on the problem at hand, a common issue facing all of them is the determination of the material-specific effective interaction parameters like the Hubbard $U$ and Hunds coupling $J$ for the correlated subspace. The lack of a reliable prediction procedure is one of the primary reasons this method cannot yet be considered truly \emph{ab initio}.

This well-known problem was pointed out soon after the introduction of the Hubbard model and early attempts to estimate the Hubbard $U$ in real materials were made by Cox et.al \cite{IOP_HubbU_transition_metal_1973_B_N_Cox}. Subsequent advances led to development of a method based on the Local Density Approximation(LDA) called cLDA (constrained LDA),  in which the Hubbard $U$ is calculated from the energy difference between different occupations of the localized orbitals after cutting off hoppings between the correlated orbitals and the itinerant valence states\cite{PRB_cLDA_Cuprates_1988_McMahan,PRB_cLDA_cuprates_1989_Christensen,PRB_cLDA_Fe_Ce_1991_Anisirnov}.
However, this method tends to overestimate $U$ since a lot of physical screening channels are eliminated when the hoppings are cut off. 
Recently, another  approach based on the Random Phase Approximation(RPA) called constrained RPA (cRPA)\cite{PRB_CRPAMethodInvented_1998_F.Ary,PRB_lowenegymodel_for_firstprincilpes__2004_F.Ary} has  gained popularity as it is material-specific and provides a clear picture of the physical screening channels which are taken into account.  cRPA has been applied to a variety of strongly correlated systems such as transition metals and their oxides\cite{PRB_CalculationofHubU_cRPA._2006_F.Ary,PRB_cRPAonTransitionMetal_2008_F.ARy,PRB_cRPAonTransition_oxide_2012_S.Biermann,PRB_cRPAonTransition_oxide_2012_P.H.Zhang,PRB_cRPAonTrans_oxides_2013_F.Ary}, early lanthanides\cite{PRB_cRPAonearlyLanthanide_2013_F.Ary,PRB_Screened_Coul_inter_cal_2014_F.Bruneval} and high-Tc superconductors\cite{JPSJ_lowenergyOFironbaseSC_2010_M.Imada,PRB_DyanmicScreening_LaCuO_2015_P.Werner_F.Ary}. However, the Hubbard $U$ predicted by cRPA is generally not in good agreement with the value required by DMFT impurity solvers to achieve agreement with experiment. One notable example \cite{PRB_Screened_Coul_inter_cal_2014_F.Bruneval} is elemental Cerium for which the $U$ predicted by cRPA is about 1$\sim$3eV, which is far smaller than the value of around 6eV used in practice. 
This is not surprising since the Ce $f$ orbital is more localized than the transition elements' $d$ orbital and cRPA is suspected to be inadequate for such strongly correlated systems. In spite of this, there has been little theoretical investigation into exactly why cRPA fails in the strongly correlated regime. Instead  most of the recent research on cRPA has focused on the energy window to be used in the cRPA procedure and the definition of the many-body model using the effective  $U$ predicted by cRPA \cite{PRB_cRPAonTrans_oxides_2013_F.Ary,JPSJ_lowenergyOFironbaseSC_2010_M.Imada}. In view of the above, we firmly believe that further investigation is required into the root causes of the failure of this method when strong correlations are present. 

In this paper,we investigate the accuracy of cRPA using a class of model Hamiltonians based on models used to study strongly correlated materials. This allows us to  study the fundamental causes for the failure of cRPA in strongly correlated systems in general, instead of merely making predictions about a specific compound. 
In all of our models, we include strong hybridization between localized and itinerant bands as the accuracy of cRPA is particularly questionable in such systems. Our models are two dimensional and we retain all density-density Hubbard interactions, reminiscent of the models used to study typical transitional metal oxides. 
We use DMFT to compute the spectra and quasiparticle residues of both the full multi-orbital model as well the effective one-orbital model using parameters obtained from cRPA. We show that \textrm{i)} cRPA has a tendency to systematically overestimate screening in the system. \textrm{ii)} We also find that for a large range of parameters, inter-orbital and weakly correlated orbitals' U parameters have little effect on the spectrum, thus negating the fundamental screening mechanisms used in cRPA. \textrm{iii}) Instead, we study a far more accurate form of W and U using the DMFT local Polarization bubble which exactly includes all local interactions. Using this new method, we show that the true screening is far less than predicted by cRPA/RPA and that the actual U predicted by this method has little frequency dependence. \textrm{iv)} We also study the fully screened interaction(W) evaluated using RPA and our new method and show that the RPA W is unable to capture the Mott transition and also shows no signatures of local screening processes, which are present in in the W evaluated using our new method. Since local interactions are treated exactly in DMFT, this success of the local Polarization method clearly shows that DMFT takes into account all the predominant screening processes in strongly correlated systems which are missing from RPA-based approaches. 


\section{Models and Methods}
\subsection{Model Hamiltonians}
We start by introducing the two models, which we name dp model (for the two-band model) and dps model(for the three-band model). For the dp model, we parametrize the  tight-binding part of our Hamiltonian using a two-component field $\psi^\dagger_{\mathbf k\sigma}=[d^\dagger_\sigma(\mathbf k),p^\dagger_\sigma(\mathbf k)]$ in which $d^\dagger_\sigma(\mathbf k)$ [$p^\dagger_\sigma(\mathbf k)$] creates a d (p) electron with spin $\sigma$ and wave vector $\mathbf k$. The Hamiltonian is given by : 
\begin{equation}
H^{dp}_0=\sum_{\mathbf{k} \sigma}\psi_{\mathbf k\sigma}^\dagger
 \left(
 \begin{array}{cc}
 \epsilon_d(\mathbf k)-\mu&t_{dp}(\mathbf k)\\
t_{dp}(\mathbf k)&\epsilon_p(\mathbf k)-\mu
 \end{array}
 \right)
\psi_{\mathbf k \sigma}
\end{equation}
where 
\begin{align*}
\epsilon_{m}(\mathbf k)    & =E_m+t_{mm}\big(\cos(k_x)+\cos(k_y)\big)\quad m\in\{p,d\}\\ 
t_{dp}(\mathbf k) & =t_{dp}\big(\sin(k_x)+\sin(k_y)\big) 
\end{align*}
This parametrization is motivated by recent research \cite{NJOP_Udp_in_cuprates_2014_K.Held} investigating the significance of $U_{dp}$ on the opening of the gap for the undoped cuprates and it describes electrons hopping on a two-dimensional lattice with two orbitals per site.The band dispersion and the one-electron thermal non-interacting Green's function matrix are given by:
\begin{align}
E_{\pm}(\mathbf k)=\epsilon_{+}(\mathbf k)\pm\sqrt{\epsilon^2_-(\mathbf k)+t^2_{dp}(\mathbf k)}-\mu\\
\hat G(k)=\frac{[i\omega_n+\mu-\epsilon_+(\mathbf k)]\hat 1+t_{dp}(\mathbf k)\hat \tau_1+\epsilon_-(\mathbf k)\hat \tau_3}{[i\omega_n-E_-(\mathbf k)][i\omega_n-E_+(\mathbf k)]}
\end{align}
where $\hat \tau_i$ denotes Pauli matrices and $\epsilon_{\pm}(\mathbf k) =\big(\epsilon_d(\mathbf k)\pm\epsilon_p(\mathbf k)\big)/2$. 

\begin{figure}[h]
 \includegraphics[width=\columnwidth]{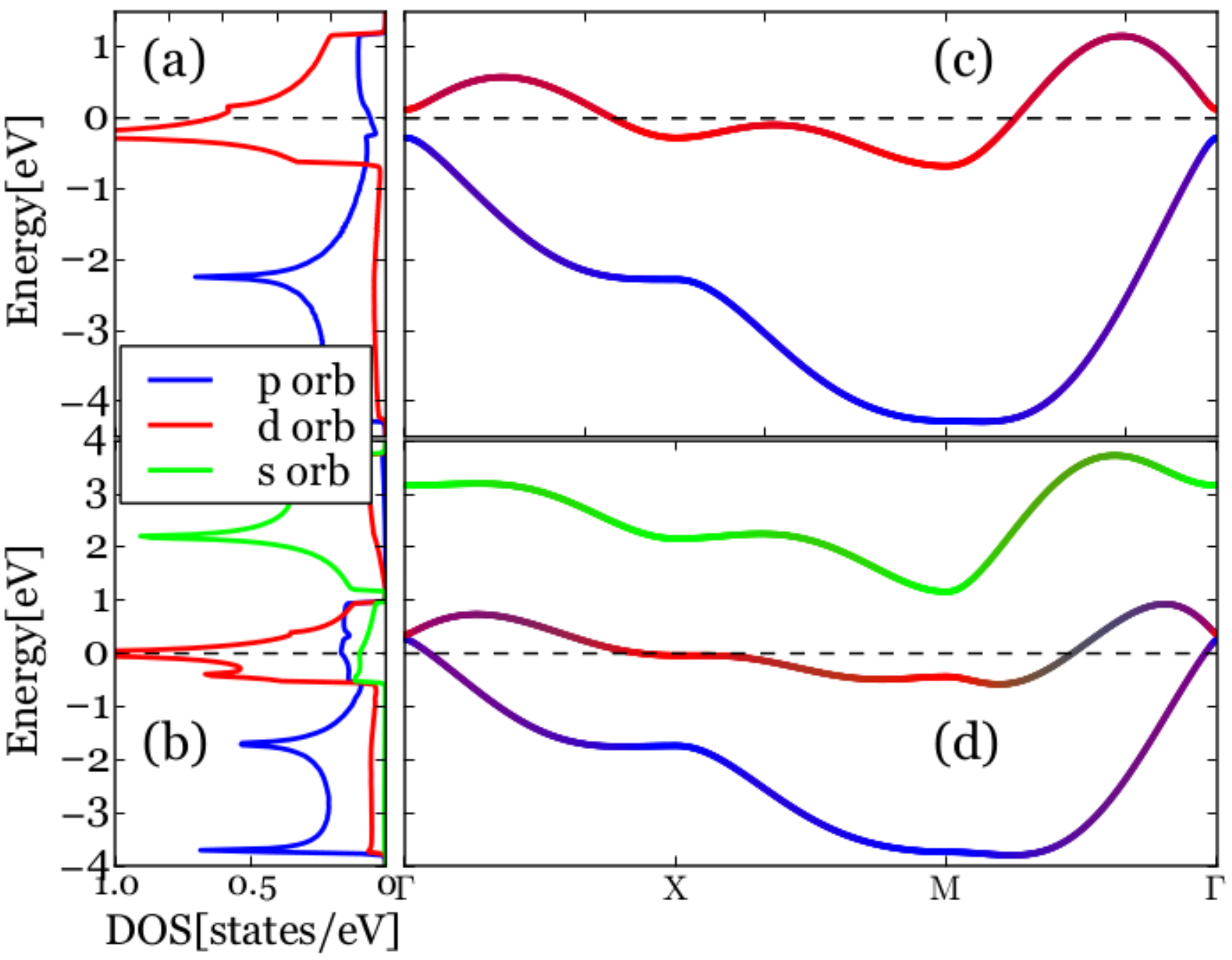}
 \caption{\label{fig1}Orbital-resolved density of states (DOS) of (a) dp model ($\mu$=0.28, $n_p$=1.78, $n_d$=1.22) and (b) dps model ($\mu$=0.038, $n_p$=1.70, $n_s$=0.14 and $n_d$=1.16). Momentum-resolved spectral functions of (c) dp model and (d) dps model along $\Gamma-X-M-\Gamma$. The coded color represents contribution from different orbitals.}
\end{figure}

In Fig.~\ref{fig1}(a) and (c) we show the non-interacting density of states(DOS) and band structure of the model with the parameters $E_p=-2.0,E_d=0.0,t_{dp}=1.0,t_{dd}=0.2$ (in units of $t_{pp}$).Unless specified otherwise,all the calculations in this paper have been performed at a fixed  total electron number per site $n=3$ and at an inverse temperature of $\beta=100$.We note that there are several Van-Hove singularities in the DOS due to the extrema in the energy spectrum. We can also see from the from the orbital-resolved DOS that there is some mixture of d and p states around the chemical potential.

Next we turn to the dps model, in which we add a third band to the dp model with the aim of enhancing the particle-hole screening excitations in the system.The tight-binding part of the dps model could be written  using a three-component field  $\phi^\dagger_{\mathbf k\sigma}=[d^\dagger_\sigma(\mathbf k),p^\dagger_\sigma(\mathbf k),s^\dagger_\sigma(\mathbf k)]$:
 
 \begin{equation}
 H^{dps}_0=\sum_{\mathbf k \sigma}\phi^\dagger_{\mathbf k\sigma}
 \left(
 \begin{array}{ccc}
 \epsilon_d(\mathbf k)-\mu&t_{dp}(\mathbf k)&t_{ds}(\mathbf k)\\
t_{dp}(\mathbf k)&\epsilon_p(\mathbf k)-\mu&0\\
t_{ds}(\mathbf k)&0&\epsilon_s(\mathbf k)-\mu
 \end{array}
 \right)
 \phi_{\mathbf k\sigma}
 \end{equation}

 For simplicity,the parameterization used  is similar to dp model 

 \begin{align*}
 \epsilon_{\alpha}(\mathbf k) &=E_\alpha+t_{\alpha\alpha}\big(\cos(k_x)+\cos(k_y)\big)\quad \alpha\in\{p,d,s\},\\
 t_{dp}(\mathbf k) &=t_{ds}(\mathbf k)=t_{dp}\big(\sin(k_x)+\sin(k_y)\big).
 \end{align*}

 Here we choose $t_{ps}(\mathbf k)=0$ with two considerations in mind: i)Physically,it is reasonable to take it to be zero as these two bands are well-separated in energy; ii) It is helpful in reducing the sign problem in our CTQMC impurity solver used while solving the DMFT equations,which are discussed in detail in section~\ref{DMFT}.\\
 Fig.~\ref{fig1} (b) and (d) show the calculated DOS and band structure of the dps model with $E_p=-1.7,E_d=0.0,E_s=2.2,t_{dd}=0.2,t_{ss}=0.5,t_{dp}=1.0$ (in units of $t_{pp}$). The basic structure of DOS resembles that of dp model except that there are more Van-Hove singularities in the dps model and more appreciable mixture of d and p,s states around the chemical potential.

For the interacting part of the Hamiltonian, we only retain all possible on-site density-density interactions and ignore exchange interactions such as Hunds Coupling. The interaction Hamiltonian is given by:
 \begin{equation}\label{int}
 H^{dp(s)}_U=\sum_{i}\bigg(\sum_{m}U_{mm}\hat{n}_{im\uparrow}\hat{n}_{im\downarrow}+\frac{1}{2}\sum_{m\neq o}U_{mo}\hat{n}_{im}\hat{n}_{io}\bigg)
 \end{equation}
 Here $i$ labels the lattice site, $m,o\in\{p,d,(s)\}$ , $\{U_{dd},U_{pp},U_{ss}\}$ represent the intra-orbital interaction strengths and  $\{U_{dp},U_{ds},U_{ps}\}$ the inter-obital interaction strengths.As the d band is taken to be the most correlated one,we place the added constraint that $U_{dd}$ is the greatest of all the U parameters.
 
\subsection{constrained Random Phase Approximation(cRPA)}
\begin{figure}[h]
 \includegraphics[width=\columnwidth, height=0.8in]{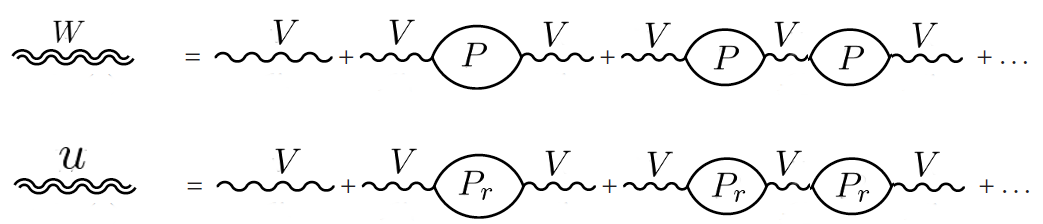}
 \caption{\label{fig2} a) Diagrams showing the RPA screening process to obtain the fully screened interaction W from the unscreened interaction V. b) Diagramss showing the cRPA process to obtain the partially screened $u$ from the unscreened interaction V.  }
\end{figure}

In this section, we describe the cRPA scheme used in this paper. In cRPA\cite{PRB_lowenegymodel_for_firstprincilpes__2004_F.Ary},the Hubbard $u^{cRPA}$ of the effective model is obtained by factoring in screening by the degrees of freedom involving the itinerant bands in an RPA-like fashion. We rewrite the total polarization function $P$ as $P=P_r+P_d$, where $P_d$ is the polarization function within the d subspace and $P_r$  contains all other terms. Using this definition, the effective $u^{cRPA}$ can be written as:
\begin{equation}
u^{cRPA}(q)=V(\mathbb{1}-P_r(q)V)^{-1} \label{cRPA Eq}
\end{equation}
with $P_r$ approximated by only the particle-hole(RPA)``bubble" diagrams.The fully screened interaction $W^{RPA}$ can be also evaluated using RPA by factoring in the screening effect of $P_d$ on $u$
\begin{equation}\label{W_def}
W^{RPA}(q)=u^{cRPA}(q)(\mathbb{1}-P_d(q)u^{cRPA}(q))^{-1}
\end{equation} 
The Feynman diagrammatic  illustration for the summation procedure of the bubble diagrams used to calculate $W^{RPA}$ or $u^{cRPA}$ is shown in Fig.~\ref{fig2}.
Since we keep only density-density interactions in the model, the polarization function $P$ in orbital basis depends on two orbital indices instead of four in general. The bubble polarization function is given by
\begin{multline}
P_{mn}(\mathbf q,\omega)=\\
\sum_{\mathbf k,\lambda,\beta}a^{m*}_{\lambda,\mathbf k}a^n_{\lambda,\mathbf k}a^{m}_{\beta,\mathbf{k-q}}a^{n*}_{\beta,\mathbf{k-q}}\frac{f(E_{\beta,\mathbf{k-q}})-f(E_{\lambda,\mathbf k})}{\omega+E_{\beta,\mathbf{k-q}}-E_{\lambda,\mathbf k}+i\delta}\label{polarzationeq}
\end{multline}
and interaction matrix $V$ in the two models are given by $(V)_{mn}=U_{mn}$ defined in Eq.~[\ref{int}],where $m,n\in\{p,d,(s)\}$ and wavefunction $a^n_{\lambda,\mathbf k}$ in $\lambda$ band is $\langle n,\mathbf k|\lambda,\mathbf k\rangle$.\\

\subsection{\label{DMFT}DMFT calculations}
In order to solve our lattice models,we employ the DMFT method which maps them onto an impurity problem subject to the following self-consistency condition\cite{RMP_DMFT_1996_A.G_G.K}:
\begin{equation}\label{dmftscc}
\sum_{\mathbf k}(i\omega\mathbb{1}-h_0(\mathbf k)+\Sigma_{DC}-\Sigma(i\omega))^{-1}=(i\omega\mathbb{1}-E_{imp}-\Sigma(i\omega)-\Delta(i\omega))^{-1}
\end{equation}
Here $h_0(\mathbf k)$ is the kernel of the two tight-binding models $H^{dp(s)}_0$, $\Delta$ is the frequency-dependent hybridization of the impurity with the bath and $\Sigma$ is the impurity Self Energy, which is approximated within DMFT to be equal to the (local) Self Energy of the system.The double counting(DC) term $\Sigma_{DC}$ is needed to subtract the part of the correlation that is overcounted  in our tight-binding models and the DMFT solution.We used the form for the DC term given in Eq.~[\ref{DC_eqn}] which generalizes the standard DC correction to multi-band systems with interorbital interactions\cite{JPCM_LDA+U_Doublecounting_1997_Anisimov}. Note that the orbital occupancies used in the equation are obtained from the solution of the non-interacting model as that accurately gives us the Hartree shifts already taken into account by our model before DMFT corrections are put in. This is also in the spirit of the Double Counting corrections usually used in LDA+DMFT calculations where the atomic occupancies are used to calculate the Double Counting. :
\begin{equation}\label{DC_eqn}
\Sigma^m_{DC}=\sum_{o\neq m}U_{mo}n_o+U_{mm}\frac{n_m}{2}
\end{equation}
The quantum impurity model is solved using the numerically exact continuous-time quantum Monte Carlo method\cite{PRL_firstpaper_on_CTQMC_2006_P.Werner_A.J,PRB_CTQMC_2007_K.Haule}. CTQMC is known to have a sign-problem when large off-diagonal terms exist in $\Delta$ or $\Sigma$. However, for the dp and dps models considered here it can be proved that most off-diagonal terms in $\Sigma$ terms vanish\footnote{We prove this using Eq.~[\ref{dmftscc}].In the case of dp model,the off-diagonal hybridization function $\Delta_{pd}\propto\sum_{\mathbf k} \frac{t_{dp}(\mathbf k)}{det(i\omega-h_0(\mathbf k)+\Sigma_{DC})} $ in the first iteration by setting $\Sigma=0$ in Eq.~[\ref{dmftscc}].This turns out to be zero because $t_{dp}(\mathbf k)$ is odd in $\mathbf k$ and the determinant is even in $\mathbf k$.This says that $\Delta$ is diagonal in the first iteration.After solving the impurity model using diagonal $\Delta$,the impurity self energy is also diagonal. Utilizing Eq.~[\ref{dmftscc}] again,one find that $\Delta$ remains diagonal for nonzero but diagonal self energy.This completes our proof that we can choose $\Delta$ and $\Sigma$ to be diagonal in our simulation.Similar analysis of dps model leads to Eq.~[\ref{selfenergy}].}.For the dp model,the self-energy matrix is exactly diagonal in the two orbital channels,while for the dps model,it is of the form:
\begin{equation}\label{selfenergy}
\Sigma=
\left(
\begin{array}{ccc}
\Sigma_{dd}&0&0\\
0&\Sigma_{pp}&\Sigma_{ps}\\
0&\Sigma_{ps}&\Sigma_{ss}
\end{array}
\right)
\end{equation}
The lattice self-energy and Green's function are obtained by iterating our equations to self-consistency. As noted earlier, the chemical potential in all our simulations is adjusted such that the total electron occupancy is  $3$. As our DMFT scheme provides us with quantities on the imaginary (matsubara) axis, in order
to obtain physical quantities on real frequency axis we use the maximum-entropy analytical continuation method\cite{PhysicsReports_MEM_1996_M.Jarrell}. Additionally, in order to estimate the degree of correlations present in our model in different simulations, we calculate the quasiparticle residue which (within the of DMFT approximation) is given by:
\begin{equation}\label{QP}
 Z_m=(1-\frac{Im\Sigma_m(i\omega)}{\omega}|_{\omega\rightarrow 0})^{-1}
\end{equation} 
where m is the band index.

\section{Results}
\subsection{\label{DOS}Density of states}
\begin{figure}[h]
 \includegraphics[width=0.9\columnwidth]{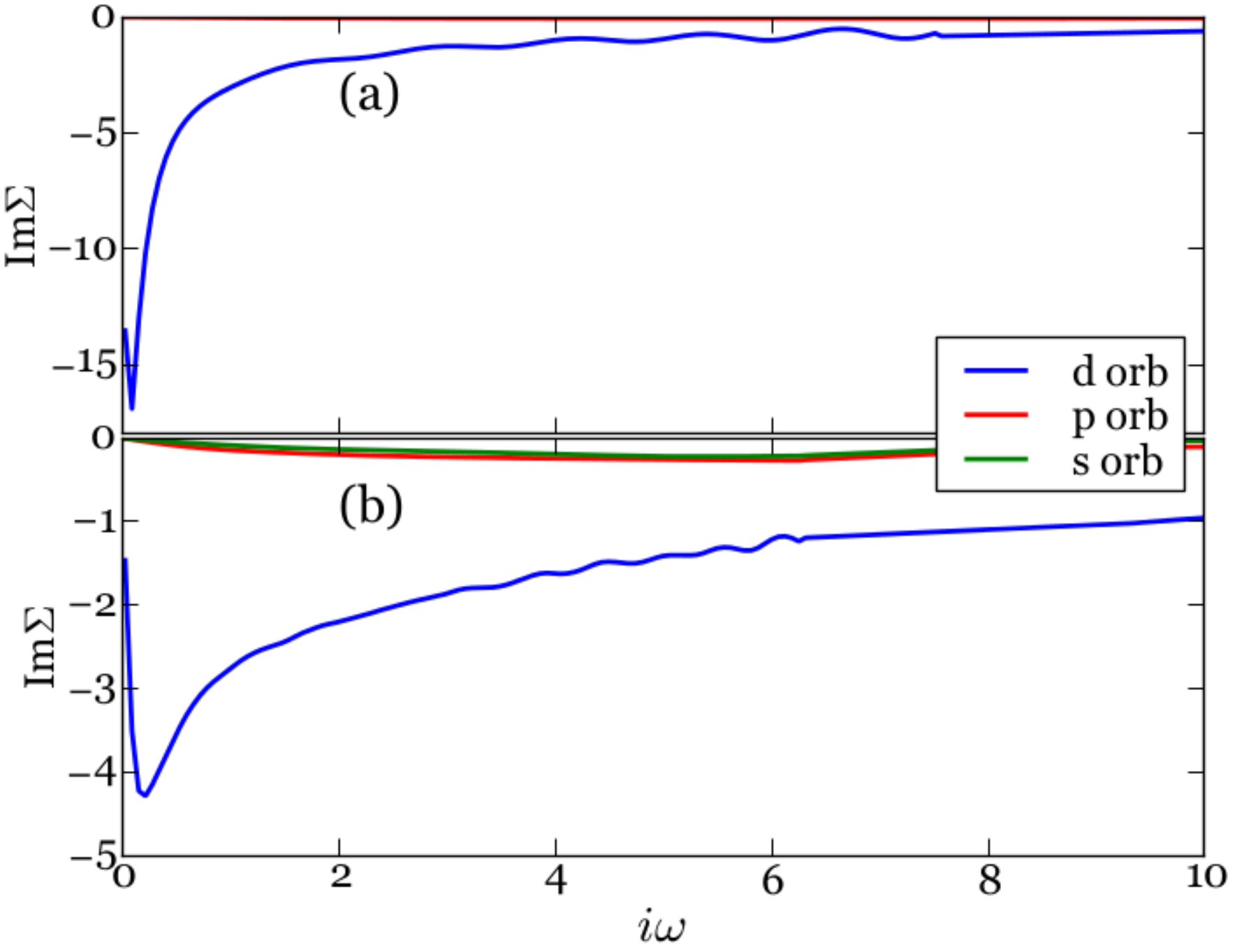}
 \caption{\label{fig3}Comparison of Im$\Sigma$ of different orbital in ``2-orb/3-orb" scenario for (a) dp model with $U_{dd}=4.5 $ and (b) dps model with $U_{dd}=6.0$.}
\end{figure}

In this section,we compare the densities of states(DOS) obtained for three different scenarios near the Metal-Mott insulator transition (MIT) for both the dp and the dps models using DMFT. The three scenarios we study are: 1) Full two-orbital(dp) or three-orbital(dps) DMFT with all on-site interactions factored in, which we  dub the ``2-orb/3-orb" scenario, 2)one orbital DMFT  where we fix the value of the effective $u_{dd}$ to the same value as full model (thereby neglecting any screening) , which we name the ``1-orb bare" scenario and 3)one orbital DMFT with an effective $u_{dd}$ on the correlated orbital calculated using cRPA,  dubbed the ``1-orb cRPA" scenario. We emphasize here that the number of bands are the same in all three scenarios considered and the differences lie in the choice of correlated orbitals and value of the interaction in these subspace.In the following, we choose the following sets of parameters for U : $(U_{pp}=0.2U_{dd},U_{pd}=0.8U_{dd})$ for the dp model and $(U_{pp}=U_{ss}=U_{ps}=0.1U_{dd},U_{pd}=0.6U_{dd},U_{ds}=0.3U_{dd})$ for the dps model. These parameters give appreciable screening by cRPA and are therefore suitable to investigate its accuracy. With these sets of U, we find that in ``2-orb/3-orb" scenario the critical U for the MIT for the dp model is $U^{MIT}_{dd}\sim 3.2$, while for the dps model  $U^{MIT}_{dd}\sim 4.5$. For the ``2-orb/3-orb" scenario,though there exist Hubbard-like interaction terms in the p or s orbitals, the self energies in these orbitals are normally negligible compared to that in the d orbital, as shown in Fig.~\ref{fig3}  for dp model with $U_{dd}=4.5$ and dps model with $U_{dd}=6.0$. This clearly shows that an effective one orbital model can be defined which reproduces the physics of the full model in both cases.

\begin{figure}[h]
 \includegraphics[width=0.9\columnwidth]{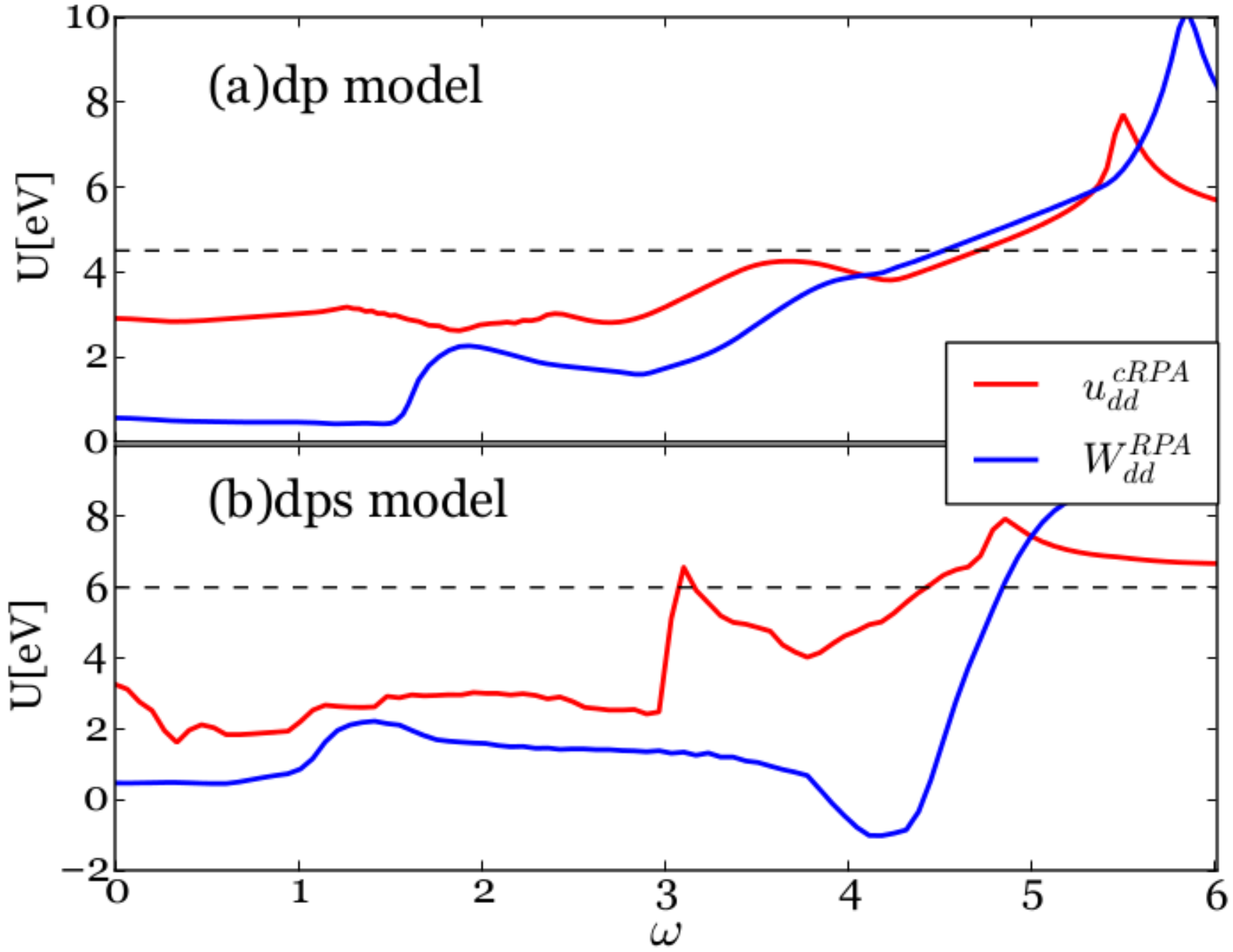}
 \caption{\label{fig4}Frequency-dependent behavior of $u^{cRPA}_{dd}$ and $W^{RPA}_{dd}$ of the (a) dp model ($U_{dd}$=4.5) and (b) dps model ($U_{dd}$=6.0) predicted by cRPA method. The dashed horizontal line denotes the bare value of $U_{dd}$ used in these two models. It is checked that both $u^{cRPA}_{dd}$ and $W^{RPA}_{dd}$ approach the bare value in the limit of high frequency as expected(not shown here). }
\end{figure}

In order to calculate the effective interaction parameters predicted by cRPA, we obtained  the screened frequency-dependent $u$ and $W$ for the critical values of $U_{dd}$ for the two models given earlier. As shown in Fig.~\ref{fig4}, cRPA predicts a static value of $u^{cRPA}_{dd}(\omega=0)= 2.91$ for the dp model and $u^{cRPA}_{dd}(\omega=0)=3.26$ for the dps model.These correspond to about $35.3\% $ and $45.7\%$ screening for the dp and dps models respectively. We also note that within an energy window $0-3eV$, $u$ is almost flat in both models, which means effective u will be very close to the static value even if one adopts a scheme accounting for frequency dependent $u$ within a finite energy window. Therefore we believe that our static $u$ based DMFT is more than adequate for these calculations.


\begin{figure}[h]
 \includegraphics[width=0.9\columnwidth]{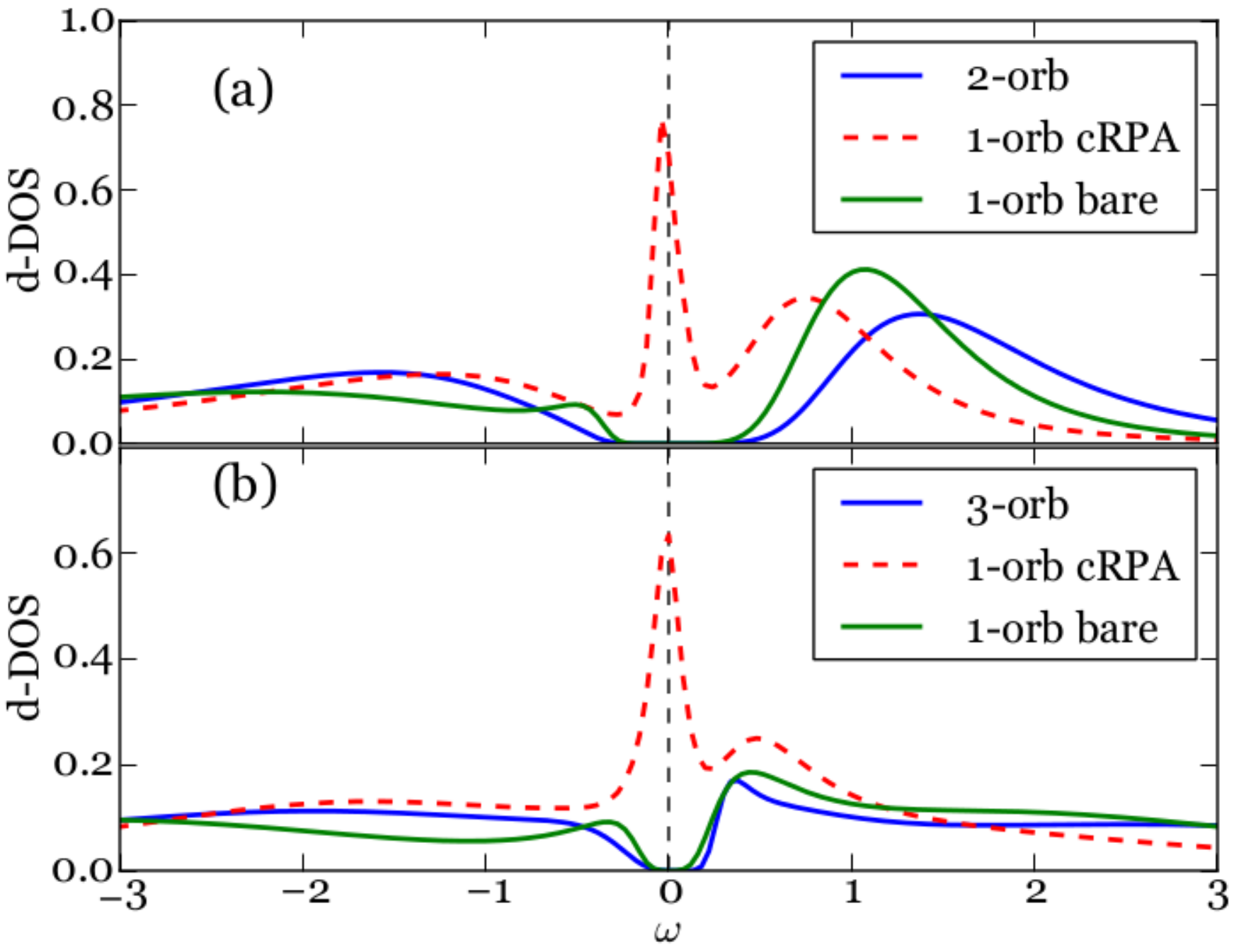}
 \caption{\label{fig5}Comparison of DOS of d orbital (d-DOS) within three scenarios for (a) dp model ($U_{dd}$=4.5) and (b) dps model ($U_{dd}$=6.0).}
\end{figure}

Next we present the central result in this section: comparing the spectral functions of the three scenarios for both models .We compare only the d-orbital DOS  as the dos of other orbitals share similar trend. As shown in Fig.~\ref{fig5},we find that for both dp and dps models with the critical parameters defined above, the ``1-orb cRPA" scenario is metallic,in sharp contrast to the Mott-insulating ``2-orb/3-orb" and ``1-orb bare" scenarios. This shows that in the models we considered the bare scenario is a much better approximation to the original many-orbital model compared to cRPA . cRPA grossly overestimates the amount of screening that is present in these models near MIT. The fact that the ``1-orb bare" scenario accurately reproduces the spectra also shows that $U_{dp}$ and $U_{pp}$ treated dynamically have very little effect on the system, which further negates the fundamental screening mechanisms used in cRPA.



\subsection{Quasiparticle Residue}
\begin{figure}[h]
 \includegraphics[width=0.9\columnwidth]{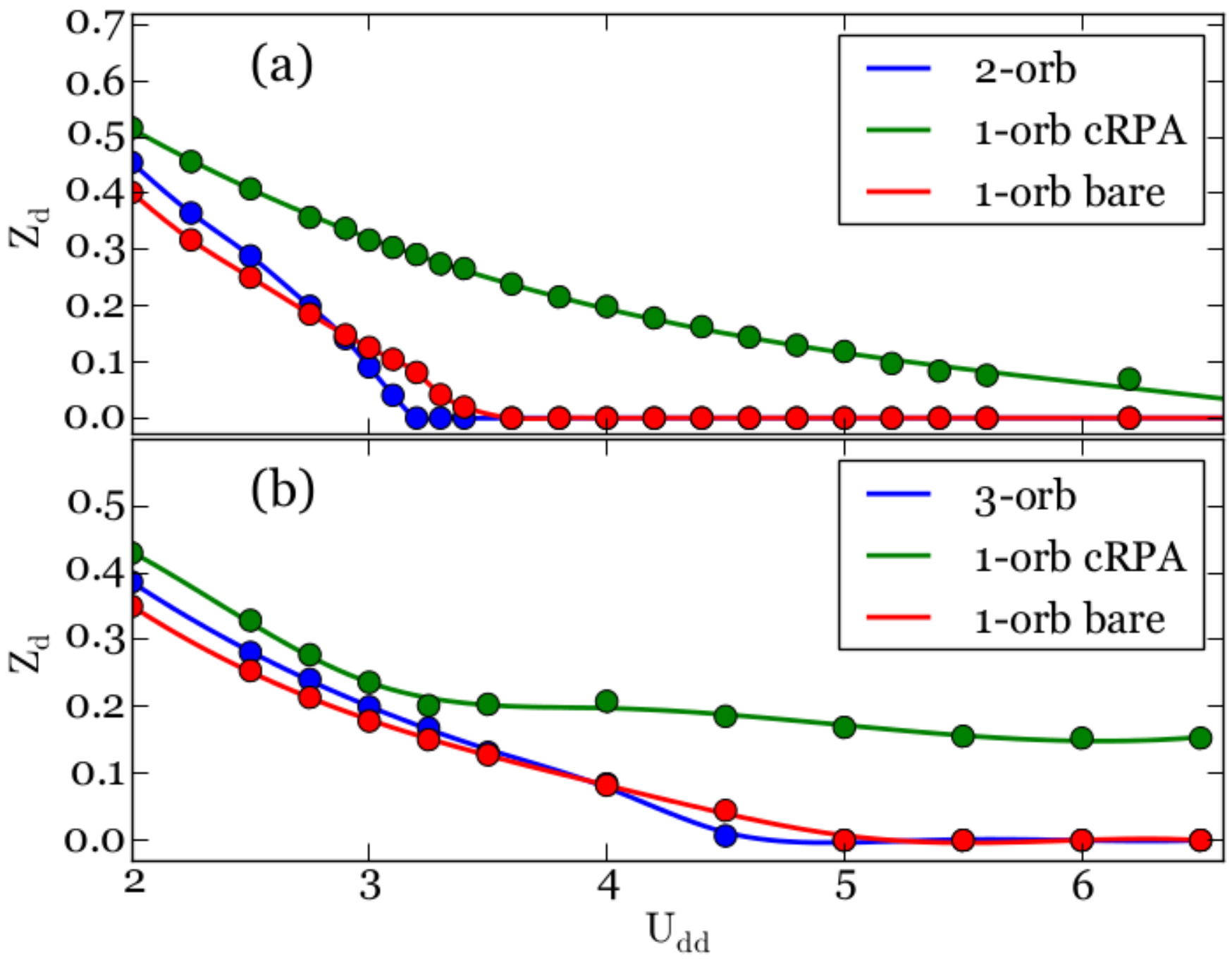}
 \caption{\label{fig6}Evolution of quasiparticle residue within the three scenarios by varying $U_{dd} $ in (a) dp model with $U_{pp}=0.2U_{dd},U_{pd}=0.8U_{dd}$ and (b) dps model with $U_{pp}=U_{ss}=U_{ps}=0.1U_{dd},U_{pd}=0.6U_{dd},U_{ds}=0.3U_{dd}$.}
\end{figure}

\begin{figure}[h]
 \includegraphics[width=0.9\columnwidth]{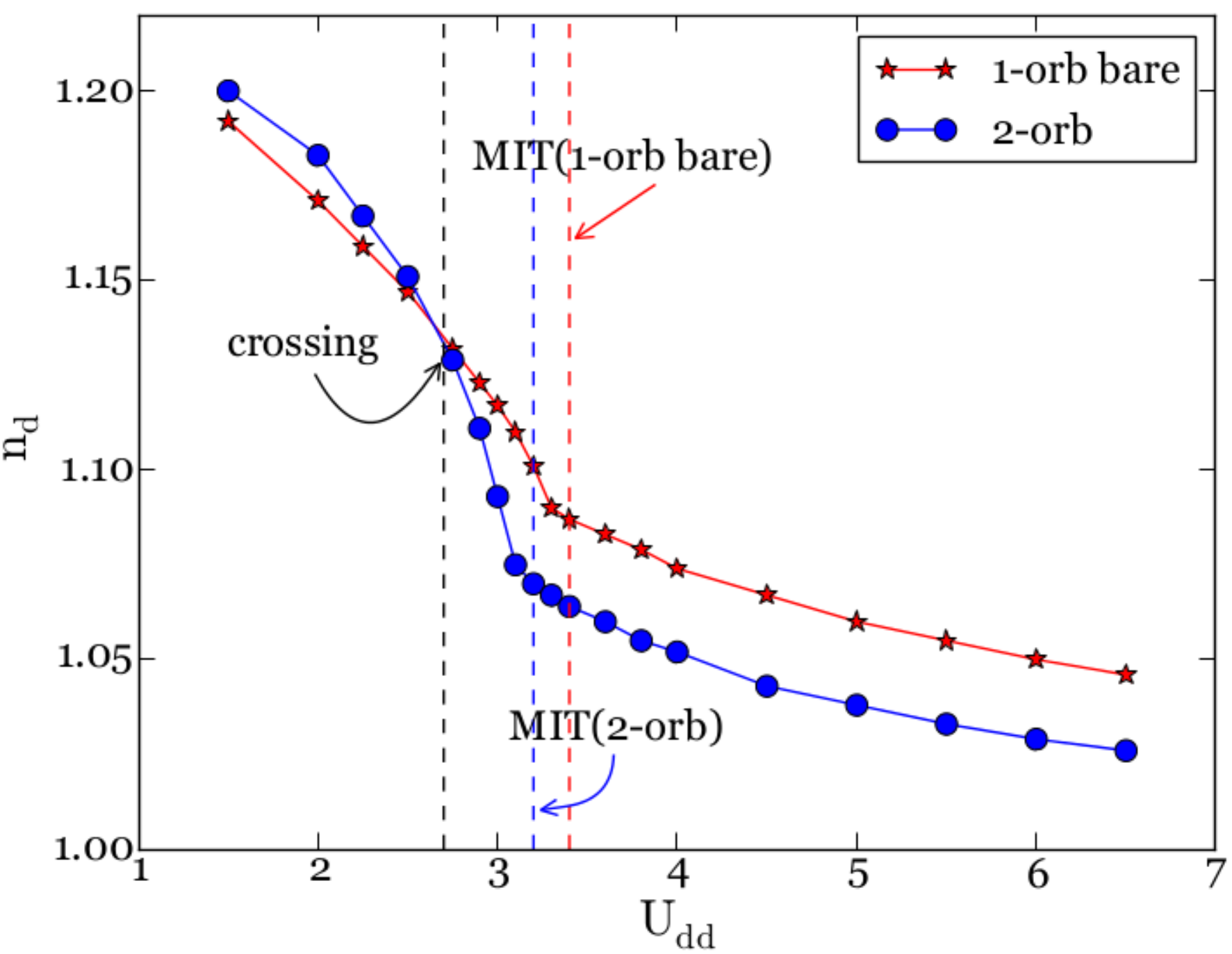}
 \caption{\label{figlast}Occupation of d orbital ($n_d$) in dp model for different value of $U_{dd}$ with $U_{pp}=0.2U_{dd},U_{pd}=0.8U_{dd}$ in ``1-orb bare" and ``2-orb" scenarios. MIT(1-orb bare) and MIT(2-orb) are transition points inferred from the quasiparticle residue results.}
\end{figure}

In order to further illustrate the inaccuracy of cRPA, we shall now show how the ``1-orb cRPA" scenario deviates from the other two for a broad range of $U_{dd}$ by comparing the quasiparticle residue $Z_d$ of the d-band (calculated using Eq.~[\ref{QP}]) in the three scenarios for both models. $Z_d$ gives us the extent of the correlations present in the correlated band and (within the DMFT approximation) is the inverse of the effective mass of the quasiparticle excitations. So a value of 1 would denote lack of correlation, whereas $Z_d \sim 0$ would signal proximity to an insulating solution with $Z_d$ becoming zero at the critical $U$. From the results shown in Fig.~\ref{fig6}, we see that cRPA always overestimates screening, with the discrepancy getting more pronounced as the $U_{dd}$ approaches the critical value for the MIT. We also notice that ``1-orb bare" is closer to the ``2-orb/3-orb" scenario than the ``1-orb cRPA" scenario and in the range $U_{dd}\sim 2.7-3.4eV$,we observe dubious antiscreening effect in ``2-orb" scenario if it is compared to``1-orb bare" scenario in dp model. We claim that this comparison is not physical in strict sense because low energy physics is not the same in these two scenarios .To demonstrate this, we investigated the behavior of occupancy of d orbital $n_d$ in these two scenarios as shown in Fig.~\ref{figlast}. One notices that there is a tiny difference of $n_d$ in these two scenarios, showing that low-energy physics in d orbital channel is not the same.Besides, the onset point of ``antiscreening" is concurrent with the  crossing point in Fig.~\ref{figlast} where $n_d$ in ``1-orb bare" starts to outweigh that in ``2-orb" scenario. From this we claim that the dubious ``antiscreening" effect is caused by the difference in $n_d$ and thus it is not a true effect here. These results establish the fact there is little screening on the most correlated orbital by the remaining orbitals in the dp and dps model and that it could be more accurate to factor in no screening at all rather than use cRPA as a predictive mechanism.  These are in agreement with those in Sec.~\ref{DOS} and suggest that in the strongly correlated regime with large hybridization between bands, the RPA bubble diagrams are not  the most relevant ones when one is describing screening by non-correlated bands. It also suggests that in such cases, we should go beyond cRPA and consider a different screening mechanism,which is discussed in Sec.~\ref{Local Screening}.

\subsection{\label{Local Screening} Estimation of Screening using Local Polarization}

\begin{figure}[h]
 \includegraphics[width=0.8\columnwidth]{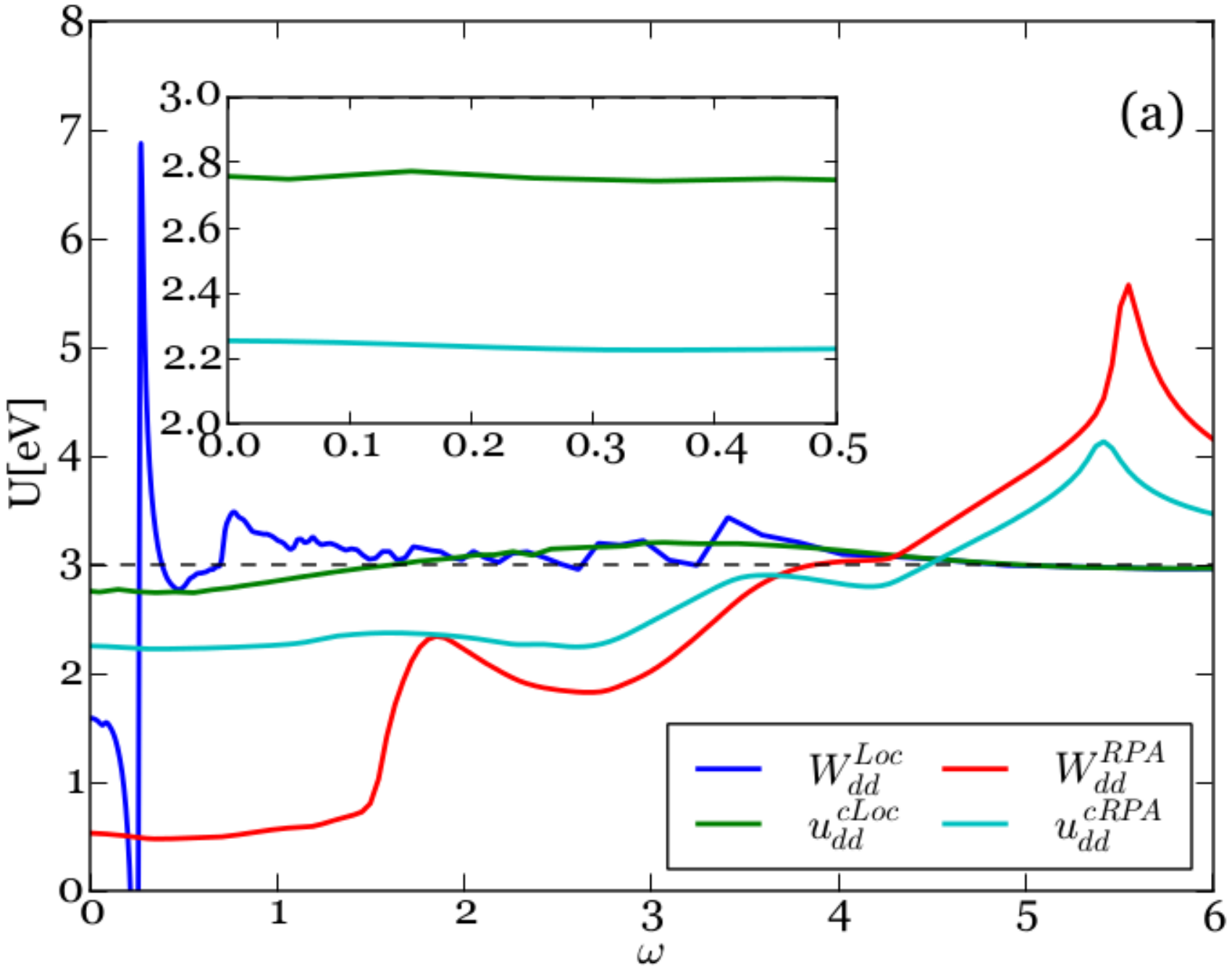}
 \includegraphics[width=0.8\columnwidth]{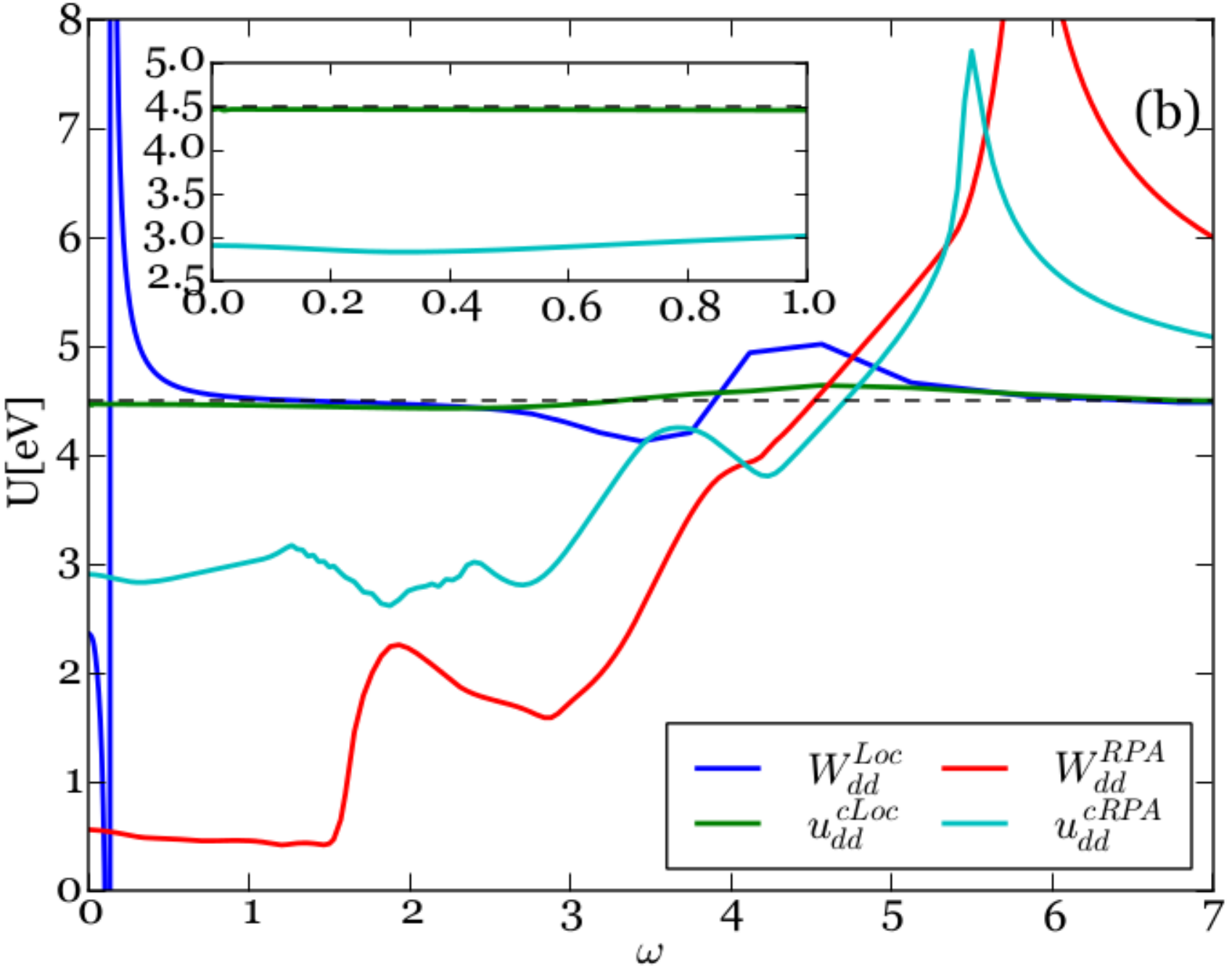}
 \caption{\label{fig7}Frequency-dependent behavior of $u^{cLoc}$ and $W^{Loc}$ with (a) $U_{dd}=3.0$ and (b) $U_{dd}=4.5$ for dp model using the Local Polarization method. The RPA $W^{RPA}$ and cRPA $u^{cRPA}$ for both sets of parameters ($U_{pp}=0.2U_{dd}, U_{dp}=0.8U_{dd}$) are also shown for comparison. Inset shows a magnified portion of the plot for $u^{cRPA}$ and $u^{cLoc}$ near $\omega \rightarrow 0$.   }
\end{figure}

We shall now explain a new formulation for estimating the screening in strongly correlated systems. As we have shown, RPA-like non-interacting bubble diagrams are inadequate in explaining the screening present in these systems. In an effort to modify this formalism and yet preserve the mathematical simplicity of the method, we shall replace the non-interacting Polarization bubble used in RPA/cRPA with the full local Polarization bubble within DMFT approximation $P^{Loc}$. Similar to cRPA, we shall also define $P^{Loc}=P_d^{Loc}+P_r^{Loc}$, where $P^{Loc}_d$ is the localized polarization in the d -subspace. We shall use this local Polarization to calculate the effective interaction in dp model using the following equations
\begin{align}\label{Uloc_def}
u^{cLoc}=V^{Loc}(\mathbb{1}- P_r^{Loc}V^{Loc})^{-1} \\
W^{Loc}=u^{cLoc}(\mathbb{1}-P_d^{Loc} u^{cLoc})^{-1}
\end{align} 
The local polarization bubbles $P^{Loc}$ here are constructed from the local two-band impurity charge and spin susceptibilities which are easily calculated using the CTQMC impurity solver \cite{Chakrabarti_Cerium}. These polarization inclusions include all the local interactions exactly and thus go far beyond the RPA-like prescription (For details on the exact procedure please refer to Appendix \ref{App.A}). Note that this new procedure also obeys the Pauli exclusion principle , which is known to be a major failing of cRPA. \cite{PRB_AccurayofcRPA_2014_P.Werner}. Also note that $u^{cLoc}$ and $W^{Loc}$ give us the screened interaction parameters in all the orbital and spin channels separately. However as  our major interest here is  the interaction between electrons with opposite spins in the correlated (d) orbital,we shall concentrate only on this particular channel. Using this new procedure we estimate the new screened interaction parameters $W^{Loc}$ and $u^{cLoc}$ for two sets of parameters, one in the correlated metallic regime with $U_{dd}=3.0$ and one in the Mott insulating regime with $U_{dd}=4.5$. The results are shown in Fig.~\ref{fig7}. The comparison between the static values of $u^{cRPA}_{dd}$ and $u^{cLoc}_{dd}$ is given in Table~\ref{table1}. As we can see, cRPA predicts vastly larger screening compared to our method (3 times larger in the metallic case and 52 times larger in the insulating case).
We performed single orbital DMFT runs using the values predicted by our new method. We see that for $U_{dd}=3.0$, our method still predicts slightly too much screening as evidenced by the enhanced metallicity of the "1-orb cLoc" run compared to the full 2-orb run. However the result is a large improvement on the cRPA result. For the insulating case, our method successfully reproduces the Mott transition, which cRPA fails completely in achieving. Our method yields a value of screened $u_{dd}$ which is almost identical to the bare $U_{dd}$, again showing that there is very little inter-orbital screening near the MIT . We see that $u^{cLoc}_{dd}$ has very little frequency dependence, which also negates the need for inclusion of more complicated frequency dependent $u$ in our impurity solvers.

\begin{table}
\caption{\label{table1}Comparison between static values of $u^{cRPA}_{dd}$ and $u^{cLoc}_{dd}$ for different values of bare U in the dp model.}
\begin{ruledtabular}
\begin{tabular}{llllll}
  bare $U$ & $u^{cRPA}_{dd}$&screening($|1-\frac{u}{U}|$)&$u^{cLoc}_{dd}$&screening\\
  3.00&2.25&25.0\%&2.76&8.00\%\\
  4.50&2.91&35.3\%&4.47&0.67\%\\
\end{tabular}
\end{ruledtabular}
\end{table}

\begin{figure}[h]
 \includegraphics[width=0.9\columnwidth]{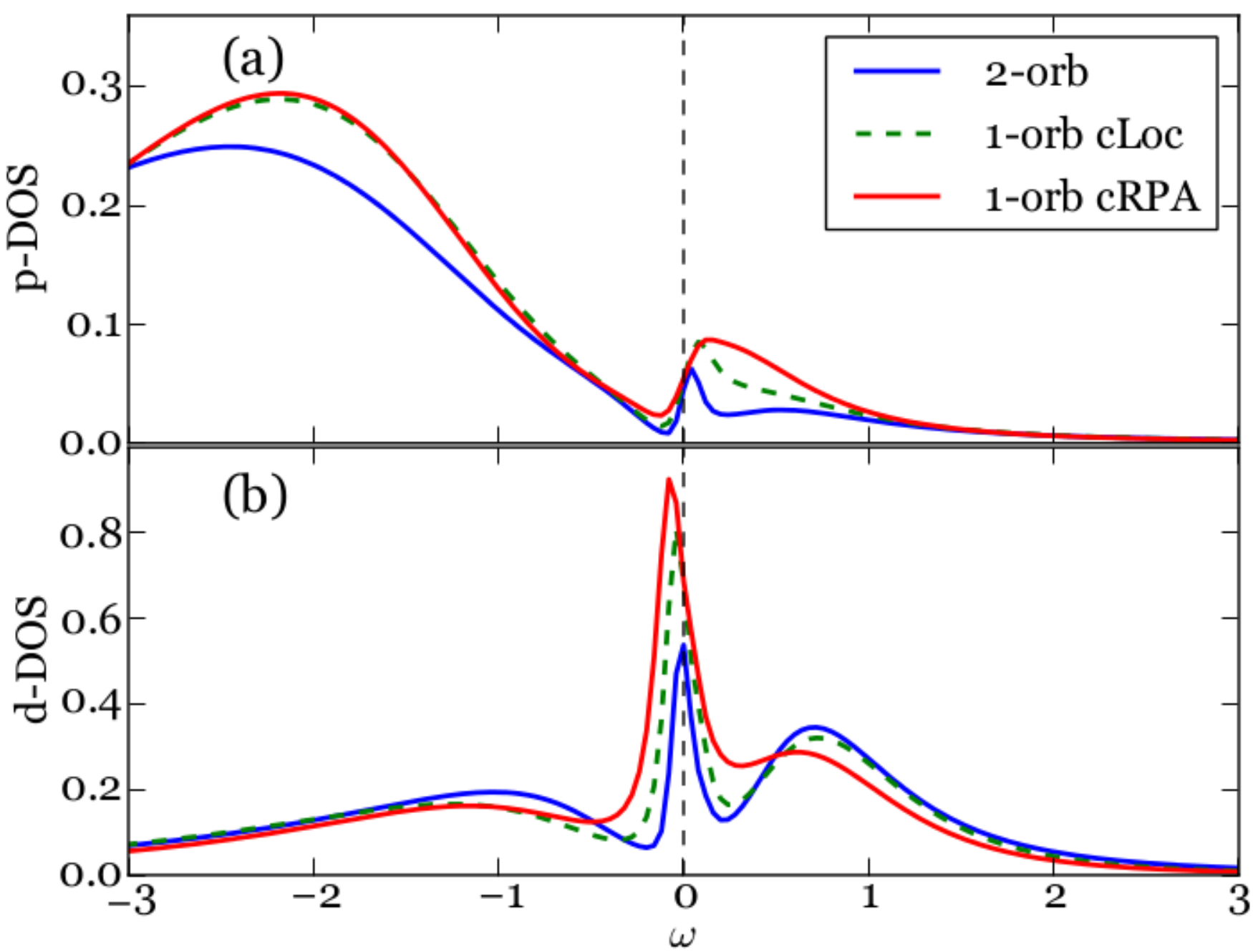}
 \includegraphics[width=0.9\columnwidth]{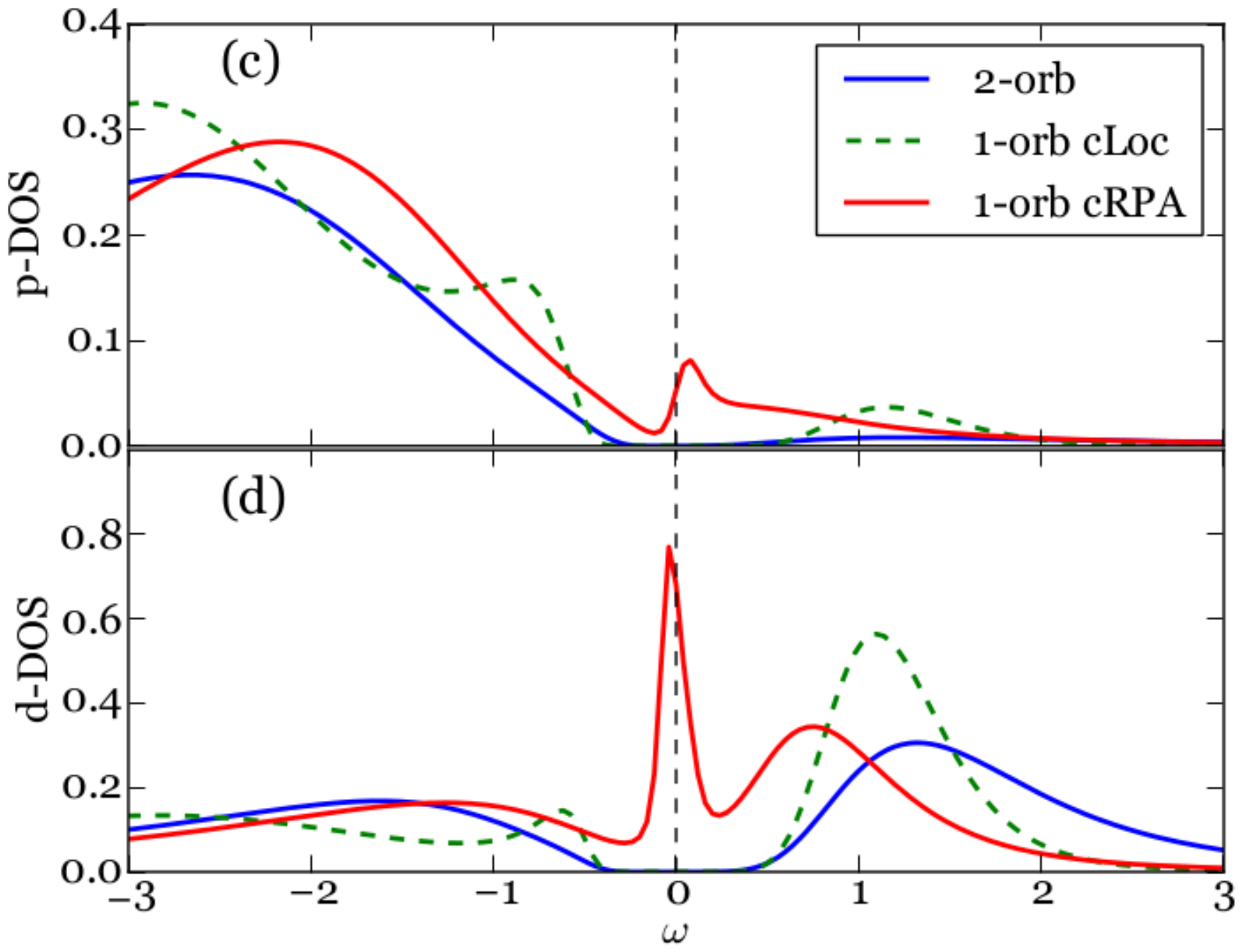}
 \caption{\label{fig8}DOS of p orbital and d orbital evaluated using DMFT within three scenarios for (a,b) $U_{dd}=3.0$ and (c,d) $U_{dd}=4.5$.}
\end{figure}
The results for the fully screened W show that RPA  had predicts a very small W $\sim 0.5$ for both the metallic and insulating cases, while our new method yields a static $W^{Loc}_{dd}$ around 2.0, thus showing much reduced screening in all channels . We also see that as we go to the insulating case, $W^{Loc}_{dd}$ displays sharper features at very low energies and has a larger static value. This pronounced low-frequency behavior is characteristic of local Kondo-like screening processes which are seen for example in the magnetic susceptibility of $\gamma$ Cerium \cite{Chakrabarti_Cerium}. Thus we see that our new prescription not only allows us to obtain values of $u$ which are much more accurate in reproducing the spectra in strongly correlated regime when using effective one-orbital models but also show clearly that local screening is by far the most dominant contribution to screening processes in these systems. It is to be noted however that our method cannot be used for prediction of effective one-orbital parameters as we would need to perform multi-orbital DMFT in order to obtain the required susceptibilities to construct the local polarization. However, this procedure clearly shows that the physics of these systems is very different from RPA and that the local inclusions to the Polarization negate some of the overscreening inherent in cRPA. Therefore we believe that any predictive method should include some treatment of local processes in order to be successful and that by calculating more accurate Polarization functions, we may be able to successfully account for screening in strongly correlated systems.

\section{Discussion and Conclusion}

In this paper,we investigate the validity of cRPA as a method for predicting the Hubbard U for strongly correlated systems.To this end we study two different models, each of which have one strongly correlated band strongly hybridized with other weakly correlated band/s.  We compare three scenarios derivable from the original lattice models and show that the full  and the ``1-orb bare" scenarios have very similar spectral functions for a wide region of interaction parameters.On the other hand,the cRPA ``1-orb cRPA" scenario with an effective on-site Hubbard $U$ calculated using cRPA has the tendency of being more metallic. We also compare the quasiparticle residue $Z_d$ of the three scenarios and show that cRPA gets progressively worse as the on-site $U_{dd}$ of the correlated band increases.These results together clearly show that cRPA has a pathological tendency  to overestimate screening in strongly correlated systems. This, we believe, is simply due to the fact that the RPA-like bubble diagrams are not the exclusive leading order terms in the screening of local Coulomb repulsion.  This systematic overscreening adds to other known deficiencies of cRPA, such as its violation of Pauli principle (note recent attempts to address this by Shinaoka et. al \cite{PRB_AccurayofcRPA_2014_P.Werner}). The extent of over-screening produced by cRPA is also hard to estimate and seems to depend on the exact dispersion of the system in question. This leads us to believe that such a mechanism is not suitable for predicting the effective screened Hubbard U for DMFT impurity solvers, except as a lower bound for any other method.  In addition, we show that in our models the interorbital interaction parameters such as $U_{dp}$ and $U_{pp}$ have very little screening effect in our model and that it is possible to stabilize a Mott Insulating phase without necessarily having large inter-orbital interactions.This is not compatible with the result obtained by Hansmann et. al \cite{NJOP_Udp_in_cuprates_2014_K.Held} in which they find a finite $U_{dp}$ is necessary to have a stable charge-transfer insulting phase. Our results seem to predict that $\epsilon_p-\epsilon_d$ and $U_{dd}$ are the important parameters which define the physics of such systems, instead of $U_{dp}$.Though our models have a different dispersion from theirs,it would be interesting to see whether treating $U_{dp}$ dynamically in their models would change their predictions. Finally we propose a new way to account for screening by using the local Polarization instead of the non-interacting Polarization bubble. We show that our new method predicts values of Hubbard U which are more accurate in reproducing the spectra of the full model, especially near the Mott Transition. Moreover the new $W^{Loc}$ also shows reduced screening compared to RPA and also indicates that highly localized Kondo-like screening in the correlated orbital is the dominant screening process, while the $W_{RPA}$ calculated using RPA fails as expected in capturing any such signatures.


\bibliography{cRPA}
\appendix
\section{\label{App.A}Evaluation of Local polarization bubbles in the dp model}
In this appendix,we sketch out the method to evaluate the partially screened $u^{cLoc}$ and fully screened $W^{Loc}$ and  elaborate how the local polarization bubbles are evaluated using impurity solver ctqmc.
Unlike the RPA method,which ignores the Pauli principle inherited in the many-body models,the formulation adopted here preserves the Pauli principle explicitly. The interaction matrix $V^{Loc}$ is :
\begin{eqnarray}
V^{Loc}=
\begin{array}{c|cccc}
&d\uparrow&d\downarrow&p\uparrow&p\downarrow\\
\hline
d\uparrow&0&U_{dd}&U_{dp}&U_{dp}\\
d\downarrow&U_{dd}&0&U_{dp}&U_{dp}\\
p\uparrow&U_{dp}&U_{dp}&0&U_{pp}\\
p\downarrow&U_{dp}&U_{dp}&U_{pp}&0
\end{array}
\end{eqnarray}
and correspondingly,the local polarization bubble matrix $P^{Loc}$ (for simplicity of notation,we denote it as $\tilde P$ afterwards)takes the form:
\begin{eqnarray}
\tilde P=
\left(
\begin{array}{cccc}
\tilde P^{d\uparrow}_{d\uparrow}&\tilde P^{d\uparrow}_{d\downarrow}&\tilde P^{d\uparrow}_{p\uparrow}&\tilde P^{d\uparrow}_{p\downarrow}\\
\tilde P^{d\downarrow}_{d\uparrow}&\tilde P^{d\downarrow}_{d\downarrow}&\tilde P^{d\downarrow}_{p\uparrow}&\tilde P^{d\downarrow}_{p\downarrow}\\
\tilde P^{p\uparrow}_{d\uparrow}&\tilde P^{p\uparrow}_{d\downarrow}&\tilde P^{p\uparrow}_{p\uparrow}&\tilde P^{p\uparrow}_{p\downarrow}\\
\tilde P^{p\downarrow}_{d\uparrow}&\tilde P^{p\downarrow}_{d\downarrow}&\tilde P^{p\downarrow}_{p\uparrow}&\tilde P^{p\downarrow}_{p\downarrow}
\end{array}
\right)
\end{eqnarray}
Because of the spin symmetry in the paramagnetic phase,we have the following identities which simplify the calculation:$\tilde P^{d\uparrow}_{d\uparrow}=\tilde P^{d\downarrow}_{d\downarrow}=a$,
$\tilde P^{p\uparrow}_{p\uparrow}=\tilde P^{p\downarrow}_{p\downarrow}=b$,$\tilde P^{d\uparrow}_{d\downarrow}=\tilde P^{d\downarrow}_{d\uparrow}=c$,$\tilde P^{p\uparrow}_{p\downarrow}=\tilde P^{p\downarrow}_{p\uparrow}=d$,$\tilde P^{d\uparrow}_{p\uparrow}=\tilde P^{d\downarrow}_{p\downarrow}=\tilde P^{p\uparrow}_{d\uparrow}=\tilde P^{p\downarrow}_{d\downarrow}=e$,$\tilde P^{d\uparrow}_{p\downarrow}=\tilde P^{d\downarrow}_{p\uparrow}=\tilde P^{p\uparrow}_{d\downarrow}=\tilde P^{p\downarrow}_{d\uparrow}=f$.To evaluate $\{a\ldots f\}$,we perform calculations of charge and spin susceptibility in different orbital channels separately.In the total orbital channel,the sampled charge susceptibility $\tilde P^{(t)}_c$ and spin susceptibility $\tilde P^{(t)}_s$ are related to $\{a\ldots f\}$ as :
\begin{eqnarray}
\tilde P^{(t)}_c=2(a+b+c+d)+4(e+f)\\ \nonumber
\tilde P^{(t)}_s=2(a+b-c-d)+4(e-f)
\end{eqnarray}
We also need to sample the charge and spin susceptibility in d and p subspace denoted as $\tilde P^{(\alpha)}_c$ and $\tilde P^{(\alpha)}_s,\alpha=p,d$ to solve the equations above.They are given in terms of $\{a\ldots f\}$ as:
\begin{eqnarray}
\tilde P^{(d)}_c=2(a+c),\tilde P^{(d)}_s=2(a-c)\nonumber\\
\tilde P^{(p)}_c=2(b+d),\tilde P^{(p)}_s=2(b-d)
\end{eqnarray}
In this way,$\{a\ldots f\}$ are given by the sampled $\tilde P^{(t)}_s,\tilde P^{(t)}_c$,$\tilde P^{(\alpha)}_c$,$\tilde P^{(\alpha)}_s(\alpha=p,d)$:
\begin{eqnarray}
a=\frac{\tilde P^{(d)}_c+\tilde P^{(d)}_s}{4},c=\frac{\tilde P^{(d)}_c-\tilde P^{(d)}_s}{4} \\
b=\frac{\tilde P^{(p)}_c+\tilde P^{(p)}_s}{4},d=\frac{\tilde P^{(p)}_c-\tilde P^{(p)}_s}{4} \\
e=\frac{1}{8}(\tilde P^{(t)}_c+\tilde P^{(t)}_s)-\frac{1}{2}(a+b)\\
f=\frac{1}{8}(\tilde P^{(t)}_c-\tilde P^{(t)}_s)-\frac{1}{2}(c+d)
\end{eqnarray}
Finally,$\tilde P_d$ is obtained by setting susceptibility in the d-d channel to be zero,i.e.$\tilde P_d=\tilde P(\tilde P^{ds}_{ds^\prime}=0),(s,s^\prime={\uparrow,\downarrow})$ and $\tilde P_r=\tilde P-\tilde P_d$.

\end{document}